\documentclass[twocolumn,nofootinbib]{revtex4}%

\usepackage{amssymb}
\usepackage{amsfonts}
\usepackage{amsmath}
\usepackage[latin1]{inputenc}
\usepackage{graphicx,color}
\usepackage{slashed}
\usepackage{young}
\usepackage[vcentermath]{youngtab}

\makeatletter

\newcommand{\be}{\begin{equation}}
\newcommand{\ee}{\end{equation}}
\newcommand{\bea}{\begin{eqnarray}}
\newcommand{\eea}{\end{eqnarray}}

\begin{document}

\title{Twisted self-duality for higher spin gauge fields and prepotentials}
\author{Marc Henneaux$^{1}$,  Sergio H\"ortner$^{2}$, Amaury Leonard$^{1}$}
\affiliation{${}^1$Universit\'e Libre de Bruxelles and International Solvay Institutes, ULB-Campus Plaine CP231, B-1050 Brussels, Belgium}
\affiliation{${}^2$Centro de Estudios Cient\'{\i}ficos (CECs), Casilla 1469, Valdivia, Chile}

\begin{abstract}
We show that the equations of motion for (free) integer higher spin gauge fields can be formulated as twisted self-duality conditions on the higher spin curvatures of the spin-$s$ field and its dual.  We focus on the case of four spacetime dimensions, but formulate our results in a manner applicable to higher spacetime dimensions.  The twisted self-duality conditions are redundant and we exhibit a non-redundant subset of conditions, which have the remarkable property to involve only first-order derivatives with respect to time.  This non-redundant subset equates the electric field of the spin-$s$ field (which we define) to the magnetic field of its dual (which we also define), and vice versa.   The non-redundant subset of twisted self-duality conditions involve the purely spatial components of the spin-$s$ field and its dual, and also the components of the fields with one zero index.  One can get rid of these gauge components by taking the curl of the equations, which does not change their physical content. In this form, the twisted self-duality conditions can be derived from a variational principle that involves prepotentials. These prepotentials are the higher spin generalizations of the prepotentials previously found in the spins 2 and 3 cases. 
The prepotentials have again the intriguing feature of possessing both higher spin diffeomorphism invariance and higher spin conformal geometry.  The  tools introduced in an earlier paper for handling higher spin conformal geometry turn out to be crucial for streamlining the analysis.  In four spacetime dimensions where the electric and magnetic fields are tensor fields of the same type, the twisted self-duality conditions enjoy an $SO(2)$ electric-magnetic invariance. We explicitly show that  this symmetry is an ``off-shell symmetry" (i.e., a symmetry of the action and not just of the equations of motion).  Remarks on the extension to higher dimensions are given.

\end{abstract}
\maketitle

\section{Introduction}
\setcounter{equation}{0}

Gravitational theories exhibit fascinating ``hidden symmetries" upon dimensional reduction \cite{Ehlers,Cremmer:1979up}.  These hidden symmetries involve duality in an essential way. For instance, in the case of $11$-dimensional supergravity reduced to four dimensions, the hidden $E_7$ symmetry includes $SO(2)$ electric-magnetic duality invariance acting on the 28 abelian gauge fields present in the theory  \cite{Cremmer:1979up}.  

It has been conjectured that these hidden symmetries might actually already be present prior to dimensional reduction, although not manifestly so, and recent analysis in the light cone formalism supports this conjecture \cite{Ananth:2016abv}.  It has also been speculated that the finite-dimensional symmetries uncovered by dimensional reduction to $D \geq 3$ dimensions are actually a subset of a much larger, infinite-dimensional Kac-Moody algebra, which could be $E_{10} $  or $E_{11}$ \cite{Julia:1980gr,Julia1,Julia:1982gx,Nicolai:1991kx,Julia:1997cy,West:2001as,Damour:2002cu} for (an appropriate extension of) maximal supergravity. 

A characteristic feature of the nonlinear realizations of the conjectured hidden symmetry algebras is that they treat the $p$-forms and their duals democratically: for each $p$-form appearing in the spectrum, the dual $D-p-2$ form also appears.  In order to exhibit the hidden symmetries of gravitational theories, it appears therefore necessary to reformulate the equations of motion for the $p$-forms present in the model in a manner that involves both the $p$-forms and their duals on an equal footing, but without doubling the number of degrees of freedom.  This is achieved
by rewriting the equations of motion as ``twisted self-duality conditions" \cite{Cremmer:1979up,Cremmer:1997ct,Cremmer:1998px}.  These conditions are first-order with respect to time (and space) and equate the electric field (respectively the magnetic field) of the $p$-form  to the magnetic field (respectively, $\pm$ the electric field) of its dual. It is easily verified that these conditions are equivalent to the Maxwell equations.  Furthermore, they can readily accomodate Chern-Simons couplings.  This form of the equations of motion for the $3$- and $7$-forms of 11-dimensional supergravity is the starting point of the authors of \cite{Tumanov:2015yjd} in their recent construction of $E_{11}$-invariant equations of motion.

The twisted self-duality conditions derive from a variational principle where both the $p$-form potential and its dual are treated as independent fields on an equal footing \cite{Bunster:2011qp}, which is however not manifestly spacetime covariant (for comments on this fact, see \cite{Bunster:2012hm}). 

Now, the non-linear realizations of all the conjectured infinite-dimensional hidden symmetries of gravitational theories  involve also the dual to the graviton \cite{West:2001as,Damour:2002cu}. For that reason, similar twisted-duality reformulations of gravity are desirable.  The task of rewriting the linearized gravitational field equations as twisted self-duality conditions putting the spin-2 field and its dual  on a democratic footing, in a manner derivable from a variational principle, has been achieved in \cite{Bunster:2012km,Bunster:2013oaa}.  

One of the original motivations underlying the $E_{10}$ conjecture \cite{Damour:2002cu} was its potential connection with the zero tension limit of string theory \cite{Gross:1988ue}.   This zero tension limit involves an infinite collection of massless higher spin fields. With this in mind, we continue in this paper our investigation of twisted self-duality  for higher spin gauge fields.  We consider explicitly the case of four spacetime dimensions but formulate our results in a manner applicable to higher spacetime dimensions \cite{Comment}.  

We establish in this paper a number of results concerning the twisted self-duality formulation of higher spin gauge fields.
\begin{itemize}
\item We show that the equations of motion of the free bosonic higher spin fields can indeed equivalently be written as twisted self-duality conditions on the curvatures of the spin-$s$ field and its dual (Section \ref{Twisted}). The crucial property that allows this reformulation is the demonstration given in \cite{Bekaert:2003az} that the equations of motion of the higher spins are equivalent to the vanishing of the Ricci tensor. This is not obvious when $s >2$ since the equations of motion are of second order, while the curvatures contain derivatives of the higher spin gauge field up to the order $s$.
\item The twisted self-duality conditions are highly redundant. In section \ref{3+1}, we decompose the higher spin curvatures into electric and magnetic components, and point out that the subset of the twisted self-duality conditions that expresses that the electric field  (respectively the magnetic field) of the spin-$s$ field is equal to the magnetic field (respectively, minus the electric field) of its dual, completely captures the full content of the twisted self-duality conditions. [The proof of this fact is postponed to Section \ref{Hamiltonian}.] Remarkably, this subset of the twisted self-duality conditions contains only first-order time derivatives of the fields  -- even though a generic curvature component can contain up to $s$ time derivatives.
\item In their ``electric-magnetic" form, the twisted self-duality conditions involve the spatial components of the spin $s$-field and its dual, as well as the components with one index in the time direction, i.e., equal to zero.  These components are pure gauge and can be eliminated by taking an appropriate curl.  The resulting equations are physically equivalent and shown to be derivable from a variational principle in Section \ref{VarPri}. This variational principle naturally involves prepotentials, which are introduced to take into acount the constraints on the electric and magnetic fields.  The prepotentials enjoy spin-$s$ diffeomorphism invariance and also, somewhat unexpectedly, spin-$s$ Weyl invariance.  The tools necessary to introduce the prepotentials have been developed in \cite{Henneaux:2015cda}, upon which we heavily rely.
\item  It turns out that the action principle so derived is exactly the action principle that one would obtain by starting from the Fronsdal action, going to the Hamiltonian formalism and solving the constraints through prepotentials.  This is proved in Section \ref{Hamiltonian}.  As a by-product of this result follows the completeness of the twisted self-duality conditions on the electric and magnetic fields.
\item Although the analysis is explicitly carried out in four spacetime dimensions, we expect that it should go through along parallel lines in higher dimensions where the higher spin field equations also admit a twisted self-duality reformulation.  Arguments supporting this expectation, and how the analysis would proceed, are outlined in Section \ref{Additional}.  There is, however, a feature peculiar to four dimensions (for the types of Young tableaux under consideration), namely, that the electric and magnetic fields are tensors of identical type, and the equations of motion are invariant under $SO(2)$ electric-magnetic duality rotations in the internal plane of the electric and magnetic fields.  The action expressed in terms of the prepotentials makes it explicit that this symmetry is an off-shell symmetry.  This result generalizes to higher spins the known results for spins $s=1$ \cite{Deser:1976iy}, $s=2$ \cite{Henneaux:2004jw,Julia:2005ze} and $s=3$ \cite{Henneaux:2015cda,Deser:2004xt}.  This is also discussed in Section \ref{Additional}. 
\end{itemize}
We conclude in Section \ref{Conclusions} with some comments.  Two appendices provide the technical steps necessary to derive the expression of the spin-$s$ field in terms of the prepotentials.  

This paper was announced in \cite{Henneaux:2015cda}, with the different title ``Emergent conformal geometry for higher
spins".

\section{Twisted Self-Duality for Higher Spin Gauge Fields}
\setcounter{equation}{0}
\label{Twisted}
\subsection{Standard form of the equations of motion}
In four dimensions, a massless field of spin (helicity) $s$ is described by a completely symmetric tensor $h_{\lambda_1 \lambda_2 \cdots \lambda_s}$ of rank $s$, subject to the gauge invariance 
\be
\delta h_{\lambda_1 \lambda_2 \cdots \lambda_s} =  s \partial_{(\lambda_1} \varepsilon_{\lambda_2 \cdots \lambda_s )} \label{GaugeInv}
\ee
where the gauge parameter $\varepsilon_{\lambda_2 \cdots \lambda_s }$ is completely symmetric but otherwise arbitrary.  The gauge invariant curvatures involve $s$ derivatives of the fields and read \cite{deWit:1979pe} 
\be
R_{\lambda_1 \mu_1 \lambda_2 \mu_2 \cdots \lambda_s \mu_s}[h] =  2^s \partial_{[\mu_1\vert} \partial_{ [ \mu_2\vert} \cdots \partial_{[\mu_s\vert } h_{\lambda_1] \vert  \lambda_2] \vert  \cdots \lambda_s]} 
\ee
It has Young symmetry type
$$
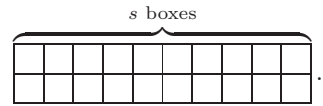
\overbrace{\yng(10,10)}^{\text{$s$ boxes}}.
$$
One can express it in terms of the derivative operator $d_{(s)}$ of \cite{DuboisViolette:1999rd,DuboisViolette:2001jk}, which fulfills $d_{(s)}^{s+1} = 0$ (see also  \cite{Coho,XBNB}).  One has $R = d_{(s)}^s h$ and $\delta h = d_{(s)} \varepsilon$ (suppressing indices), so that $\delta R=d_{(s)}^{s+1} \varepsilon = 0$.  The curvature tensor fulfills also the ``Bianchi identities"
\be
d_{(s)} R = 0.
\ee

In order to get gauge invariant objects that involve no more than second order derivatives, it is necessary to restrict the gauge parameter $\varepsilon_{\lambda_1 \cdots \lambda_{s-1}}$ to be traceless when $s>2$ \cite{Campoleoni:2012th}. The Fronsdal tensor \cite{Fronsdal:1978rb},
\bea
F_{\lambda_{1}\cdots\lambda_{s}}&=&\Box h_{\lambda_{1}\cdots\lambda_{s}} - s\partial_{(\lambda_{1}\vert}\partial^{\mu}h_{\mu\vert\lambda_{2}\cdots\lambda_{s})}\nonumber\\
&& + \frac{s(s-1)}{2}\partial_{(\lambda_{1}}\partial_{\lambda_{2}}h_{\lambda_{3}\cdots\lambda_{s})}
\eea
which involves only second order derivatives of the gauge field, is easily seen to be invariant when $\varepsilon_{\lambda_1 \cdots \lambda_{s-1}}$ is traceless.  In that approach where the gauge parameter is restricted by trace conditions, the equations of motion are actually 
\be 
F_{\lambda_{1}\cdots\lambda_{s}} =0 \label{FronsdalForm}
\ee
and are derived from the Fronsdal action \cite{Fronsdal:1978rb}
\bea
&&S[h_{\mu_1 \mu_2 \cdots \mu_s}] = \int d^{4}x[-\frac{1}{2}(\partial_{\rho}h_{\mu_{1}\cdots\mu_{s}})^{2}\nonumber\\
&&+\frac{s}{2}(\partial^{\rho}h_{\rho\mu_{1}\cdots\mu_{s-1}})^{2}-\binom{s}{2}\partial^{\rho}h_{\rho\sigma\mu_{1}\cdots\mu_{s-2}}\partial^{\sigma}\bar{h}^{\mu_{1}\cdots\mu_{s-2}}\nonumber\\
&&+\frac{1}{2}\binom{s}{2}(\partial_{\rho}\bar{h}_{\mu_{1}\cdots\mu_{s-2}})^{2}
+\frac{3}{4}\binom{s}{3}(\partial^{\rho}\bar{h}_{\rho\mu_{1}\cdots\mu_{s-3}})^2]\nonumber\\
\eea
(where $\bar{h}_{\mu_{1}\cdots\mu_{s-2}} = h^{\nu}_{\phantom{\nu}\nu\mu_{1}\cdots\mu_{s-2}}$)which is easily verified to be gauge invariant up to a total derivative (with traceless gauge parameter). 

\subsection{Equations in terms of the curvature}
In a very beautiful piece of work \cite{Bekaert:2003az}, it has been shown that the tracelessness condition on the gauge parameter is not necessary and can be viewed as a partial gauge condition.  Equivalent ideas were formulated in \cite{Francia:2002aa}, but their realization involves non-local terms, and for that reason we shall follow here \cite{Bekaert:2003az}.  

The equations of motion for a spin $s$ gauge field can be taken to be
\be
\bar{R}_{\lambda_1 \lambda_2 \lambda_3 \mu_3  \cdots \lambda_{s} \mu_{s}} = 0 \label{RicciForm}
\ee
where $\bar{R}$ is the ``Ricci" tensor obtained by taking one trace on the Riemann tensor,
\bea
\bar{R}_{\lambda_1 \lambda_2 \lambda_3 \mu_3  \cdots \lambda_{s} \mu_{s}}&=&2^{s-2}\{\Box\partial_{[\mu_{3}\vert}\cdots\partial_{[\mu_{s}\vert}h_{\lambda_{1}\lambda_{2}\vert\lambda_{3}]\cdots\lambda_{s}]}\nonumber\\ 
&& - \partial_{\lambda_{1}}\partial^{\mu}\partial_{[\mu_{3}\vert}\cdots\partial_{[\mu_{s}\vert}h_{\mu\lambda_{2}\vert \lambda_{3}]\vert\cdots\lambda_{s}]}\nonumber\\
&&- \partial_{\lambda_{2}}\partial^{\mu}\partial_{[\mu_{3}\vert}\cdots\partial_{[\mu_{s}\vert}h_{\mu\lambda_{1}\vert \lambda_{3}]\vert\cdots\lambda_{s}]}\nonumber\\
&& + \partial_{\lambda_{1}}\partial_{\lambda_{2}}\partial_{[\mu_{3}\vert}\cdots\partial_{[\mu_{s}\vert}\bar{h}_{\lambda_{3}]\vert\cdots\lambda_{s}]}\}\nonumber\\
\eea
The equations (\ref{RicciForm}) are differential equations of order $s$, but contrary to (\ref{FronsdalForm}), they are invariant under the full gauge symmetry (\ref{GaugeInv}) without restriction on the trace of the gauge parameter. It was shown in \cite{Bekaert:2003az} that they imply, with an appropriate choice of the trace of $\varepsilon_{\lambda_1 \cdots \lambda_{s-1}}$, the Fronsdal equations (\ref{FronsdalForm}) -- which, conversely, are easily verified to imply (\ref{RicciForm}).

As also pointed out in \cite{Bekaert:2003az}, the equations (\ref{RicciForm}) are very convenient for discussing duality along the lines of \cite{Hull:2001iu}.  Let $S_{\lambda_1 \mu_1 \lambda_2 \mu_2 \cdots \lambda_s \mu_s}$ be the tensor dual to $R_{\lambda_1 \mu_1 \lambda_2 \mu_2 \cdots \lambda_s \mu_s}$ on the first two indices (say),
\bea S_{\lambda_1 \mu_1 \lambda_2 \mu_2 \cdots \lambda_s \mu_s} &=& \; \!^*R_{\lambda_1 \mu_1 \lambda_2 \mu_2 \cdots \lambda_s \mu_s} \nonumber \\
&= &\frac12 \epsilon_{\lambda_1  \mu_1}^{\; \; \; \; \; \; \rho_1 \sigma_1} R_{\rho_1 \sigma_1 \lambda_2 \mu_2 \cdots \lambda_s \mu_s}
\eea
The equations of motion (\ref{RicciForm}) imply that $S_{\lambda_1 \mu_1 \lambda_2 \mu_2 \cdots \lambda_s \mu_s}$ fulfills the cyclic identity, i.e., is  a tensor of same Young type as $R_{\lambda_1 \mu_1 \lambda_2 \mu_2 \cdots \lambda_s \mu_s}$.   Furthermore, the cyclic identity for $R_{\lambda_1 \mu_1 \lambda_2 \mu_2 \cdots \lambda_s \mu_s}$ implies that $S_{\lambda_1 \mu_1 \lambda_2 \mu_2 \cdots \lambda_s \mu_s}$ is traceless, $\bar{S}_{\lambda_1 \lambda_2 \lambda_3 \mu_3 \cdots \lambda_s \mu_s} = 0$.  There is thus complete symmetry between the equations fulfilled by $R$ and its dual $S$.

\subsection{Twisted Self-Duality} 
It is this symmetry which is embodied in the twisted self-duality formulation.  When the equations of motion for the spin-$s$ field are fulfilled,  the tensor $S$ dual to the curvature not only is of same Young symmetry type 
$$
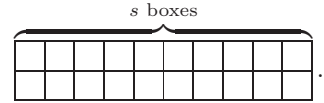
\overbrace{\yng(10,10)}^{\text{$s$ boxes}}.
$$
as $R$, but it fulfills also the Bianchi identities $d_{(s)} S=0$.  This implies the existence of a ``dual" spin-$s$ field $f_{\lambda_1 \lambda_2 \cdots \lambda_s}$ of which $S$ is the curvature \cite{DuboisViolette:1999rd,DuboisViolette:2001jk}, $$S = d_{(s)}^s f.$$  This second spin-$s$ field has its own gauge invariance since it is determined up to the $d_{(s)}$ of some $\eta_{\lambda_1 \cdots \lambda_{s-1}}$,
\be
\delta f_{\lambda_1 \lambda_2 \cdots \lambda_s} = s \partial_{(\lambda_1} \eta_{\lambda_2 \cdots \lambda_s)}
\ee

We can thus rewrite the equations of motion for the spin-$s$ theory in a duality-symmetric way where both the spin-$s$ field and its dual appear on an equal footing as follows,
\begin{equation}
{\mathcal F} = {\mathcal S}  \, ^*\hspace{-.05cm}{\mathcal F}, \label{key1}
\end{equation}
where, 
\vspace{.5cm}
\begin{equation}
{\mathcal F} = 
 \begin{pmatrix} R[h] \\ S[f] \\ \end{pmatrix}, \; \; \; ^*\hspace{-.05cm}{\mathcal F} = 
 \begin{pmatrix} ^*\hspace{-.05cm}R[h] \\ ^*\hspace{-.05cm}S[f] \\ \end{pmatrix}, \; \; \;  {\mathcal S}  = \begin{pmatrix} 0& -1 \\ 1 & 0 \end{pmatrix}  .\label{key2}
\end{equation}
This form of the equations is completely equivalent to the original form $\bar{R}=0$, since we have seen that the equation $\bar{R}=0$ implies (\ref{key1}).  And conversely, if (\ref{key1}) holds, then both  $h$ and $f$ obey $\bar{R}[h] = 0$, $\bar{S}[f] =0$, i.e., fulfill the spin-$s$ equations of motion.  Furthermore,  the two spin-$s$ fields are not independent since $f$ is completely determined by $h$ up to a gauge transformation and therefore carries no independent physical degrees of freedom.

Following \cite{Cremmer:1998px}, one refers to (\ref{key1}) as the twisted self-dual formulation of the spin-$s$ theory.

\section{Electric and magnetic fields}
\label{3+1}
\setcounter{equation}{0}
\subsection{Definitions}
The twisted self-duality conditions in their covariant form (\ref{key1}) are highly redundant.  We shall extract from them an equivalent subset that has the interesting property of containing only first order derivatives with respect to time.

To that end, we first define the electric and magnetic components of the Weyl tensor, which coincides on-shell with the Riemann tensor.  It would seem natural to define the electric components as the components of the Weyl tensor with the maximum number of indices equal to zero (namely $s$), and the magnetic components as the components with the maximum number minus one of indices equal to zero (namely $s-1$). By the tracelessness conditions of the Weyl tensor, the electric components can be related to the components with no zeroes when $s$ is even, like for gravity, or just one zero when $s$ is odd, like for Maxwell.  It turns out to be more convenient for dynamical purposes to define the electric and magnetic components starting from the other end, i.e., in terms of components with one or no zero.  Now, it would be cumbersome in the general analysis to have a definition of the electric and magnetic components that would depend on the spin.  For that reason, we shall adopt a definition which is uniform for all spins, but which coincides with the standard conventions given above only for even spins.  It makes the Schwarzschild field ``electric", but the standard electric field of electromagnetism is viewed as ``magnetic".   Since the electric (magnetic) components of the curvature of the spin-$s$ field are the magnetic (electric) components of the curvature of the dual spin-$s$ field, this is just a matter of convention, but this convention may be confusing when confronted with the standard Maxwell terminology.

Before providing definitions, we recall that the curvature $R_{i_1 j_1\cdots i_s j_s}$ of the three-dimensional ``spin-$s$ field" $h_{i_1 \cdots i_s}$ given by the spatial components  of the spacetime spin-$s$ field $h_{\lambda_1 \cdots \lambda_s}$ is completely equivalent to its Einstein tensor defined as
\be
G^{i_1 \cdots i_s} = \frac{1}{2^s} \epsilon^{i_1 j_1 k_1} \cdots \epsilon^{i_s j_s k_s} R_{j_1 k_1\cdots j_s k_s}
\ee
This tensor is completely symmetric and identically conserved,
\be
\partial_{i_1} G^{i_1 i_2 \cdots i_s} = 0
\ee
In the sequel, when we shall refer to the Einstein tensor of the spin $s$ field, we shall usually mean this three-dimensional Einstein tensor (the four-dimensional Einstein tensor vanishes on-shell).

We now define precisely the electric and magnetic fields off-shell as follows:
\begin{itemize} 
\item The electric field ${\mathcal E}^{i_1 \cdots i_s}$ of the spin-$s$ field $h_{\lambda_1 \cdots \lambda_s}$ is equal to the Einstein tensor $G^{i_1 \cdots i_s}$ of its spatial components $h_{i_1 \cdots i_s}$,
\be {\mathcal E}^{i_1 \cdots i_s} = G^{i_1 \cdots i_s}
\ee
By construction, the electric field fully captures the spatial curvature and involves only the spatial components of the spin-$s$ field.  It is completely symmetric and conserved,
\be
 {\mathcal E}^{i_1 \cdots i_s} =  {\mathcal E}^{(i_1 \cdots i_s)}, \; \; \; \partial_{i_1} {\mathcal E}^{i_1 i_2 \cdots i_s} = 0
\ee
\item The magnetic field ${\mathcal B}^{i_1 \cdots i_s}$ of the spin-$s$ field $h_{\lambda_1 \cdots \lambda_s}$ is equivalent to the components with one zero of the spacetime curvature tensor and is defined through
\be {\mathcal B}_{i_1 \cdots i_s} = \frac{1}{2^{s-1}} R_{0i_1}^{\; \; \; \; \;  j_2 k_2 \cdots j_s k_s} \epsilon_{i_2 j_2 k_2} \cdots \epsilon_{i_s j_s k_s}
\ee
It contains one time (and $s-1$ space) derivatives of the spatial components $h_{i_1 \cdots i_s}$, and $s$ derivatives of the mixed components $h_{0i_2 \cdots i_s}$.  The magnetic field is symmetric and transverse in its last $s-1$ indices,
\be
{\mathcal B}^{i_1 i_2 \cdots i_s} = {\mathcal B}^{i_1 (i_2 \cdots i_s)}, \; \; \; \partial_{i_2} {\mathcal B}^{i_1 i_2 \cdots i_s} =0.
\ee
It is also traceless on the first index and any other index,
\be
\delta_{i_1 i_2} {\mathcal B}^{i_1 i_2 \cdots i_s} = 0.
\ee
It is useful to make explicit the dependence of the magnetic field -- or equivalently, $R_{0i_1 j_2k_2 \cdots j_s k_s}$ -- on $h_{0i_2 \cdots i_s}$.   One finds
\be
R_{0i_1 j_2k_2 \cdots j_s k_s}  = \partial_{i_1} \left(d_{(s-1)}^{s-1} N \right)_{j_2 k_2 \cdots j_s k_s} + ``more" \label{ambiguity}
\ee
where ``$more$" involves only spatial derivatives of $\dot{h}_{i_1 \cdots i_s}$ and where $N_{i_1 \cdots i_{s-1}}$ stands for $h_{0i_1 \cdots i_{s-1}}$, i.e., $N_{i_1 \cdots i_{s-1}} \equiv h_{0i_1 \cdots i_{s-1}}$.
\end{itemize}

Similar definitions apply to the dual spin-$s$ field $f_{\lambda_1 \cdots \lambda_s}$.

The electric and magnetic fields possess additional properties on-shell.  First, the electric field is traceless as a result of the equation $\bar{R}^0_{\; \; 0 i_5 \cdots i_s} - \frac12 \delta^0_0 \bar{\bar{R}}_{i_5 \cdots i_s} = 0$,
\be
\delta_{i_1 i_2} {\mathcal E}^{i_1 i_2 \cdots i_s} = 0.
\ee
Second, the magnetic field is symmetric as a result of the equation $\bar{R}_{0i_4 i_5 \cdots i_s} =0$,
\be
 {\mathcal B}^{i_1 \cdots i_s} =  {\mathcal B}^{(i_1 \cdots i_s)} = 0.
\ee

We also note that there are no other independent components of the spacetime curvature tensor on-shell, since components with more than one zero can be expressed in terms of components with one or no zero through the equations of motion.

\subsection{Twisted self-duality in terms of electric and magnetic fields}
It is clear that the twisted self-duality conditions (\ref{key1}) with all indices being taken to be spatial read
\begin{equation}
 \begin{pmatrix} {\mathcal E}^{i_1 i_2 \cdots i_s}[h] \\ {\mathcal E}^{i_1 i_2 \cdots i_s}[f] \\ \end{pmatrix} = \begin{pmatrix} {\mathcal B}^{i_1 i_2 \cdots i_s}[f] \\ -{\mathcal B}^{i_1 i_2 \cdots i_s}[h] \\ \end{pmatrix} . \label{key3}
\end{equation}
It turns out that these equations are completely equivalent to the full set of twisted self-duality conditions.  This is not surprising since the components of the curvature tensor with two or more zeroes are not independent on-shell from the components with one or no zero.  The fact that (\ref{key3}) completely captures all the equations of motion will be an automatic consequence of our subsequent analysis and so we postpone its proof to later (Section \ref{Hamiltonian} below).

\subsection{Getting rid of the Lagrange multipliers}
While a generic component of the curvature may contain up to $s$ time derivatives, the twisted self-duality conditions (\ref{key3}) contain only the first-order time derivatives $\dot{h}_{i_1 \cdots i_s}$ and $\dot{f}_{i_1 \cdots i_s}$.   One can give the fields $h_{i_1 \cdots i_s}$ and $f_{i_1 \cdots i_s}$ as Cauchy data on the spacelike hypersurface $x^0 = 0$.  The subsequent values of these fields are determined by the twisted self-duality conditions up to gauge ambiguities.  The Cauchy data $h_{i_1 \cdots i_s}$ and $f_{i_1 \cdots i_s}$ cannot be taken arbitrarily but must be such that their respective electric fields are both traceless since this follows from ${\mathcal E} = \pm {\mathcal B}$ and the fact that the magnetic field is traceless.  The constraints are equivalent to the condition that the traces of the Einstein tensors of both $h$ and $f$ should be zero,
\be
\bar{G}^{i_1 \cdots i_{s-2}} [h] = 0, \; \; \; \; \bar{G}^{i_1 \cdots i_{s-2}} [f] = 0 \label{constraints0}
\ee

The twisted self-duality conditions involve also the mixed components $h_{0i_2 \cdots i_s}$ and $f_{0i_2 \cdots i_s}$.  These are pure gauge variables, which act as Lagrange multipliers for constraints in the Hamiltonian formalism.  It is useful for the subsequent discussion to get rid of them.  Since they occur only in the magnetic fields, and through a gradient, this can be achieved by simply taking a curl on the first index.    Explicitly, from the twisted self-duality conditions (\ref{key3}) rewritten as
\be
{\mathcal E}^{a \, i_1 \cdots i_s} = \epsilon^{a}_{\; \; b} \, {\mathcal B}^{b \, i_1 \cdots i_s} \label{key4}
\ee
(${\mathcal E}^{a \, i_1 \cdots i_s}  \equiv {\mathcal E}^{ i_1 \cdots i_s} [h^a]$, ${\mathcal B}^{a \, i_1 \cdots i_s}  \equiv {\mathcal B}^{ i_1 \cdots i_s} [h^a]$, $a=1,2$, $(h^a) = (h,f)$, $\epsilon_{ab} = - \epsilon_{ba}$, $\epsilon_{12} = 1$), follows obviously the equation
\be
\epsilon_{jki_1} \partial^{k} {\mathcal E}^{a \, i_1 \cdots i_s} = \epsilon^{a}_{\; \; b} \, \epsilon_{jki_1} \partial^{k}{\mathcal B}^{b \, i_1 \cdots i_s} \label{key5}
\ee
which does not involve the mixed components $h_{0i_2 \cdots i_s}$ or $f_{0i_2 \cdots i_s}$ any more.

The equations (\ref{key5}) are physically completely equivalent to (\ref{key4}).  Indeed, it follows from (\ref{key5}) that 
\be
{\mathcal E}^{a \, i_1 \cdots i_s} = \epsilon^{a}_{\; \; b} \, \tilde{ {\mathcal B}}^{b \, i_1 \cdots i_s}
\ee
where $\tilde{ {\mathcal B}}^{b \, i_1 \cdots i_s}$ differs from the true magnetic field ${\mathcal B}^{b \, i_1 \cdots i_s}$ by an arbitrary gradient in $i_1$, or, in terms of the corresponding curvature components 
\be
\tilde{R}^a_{0i_1 j_2 k_2 \cdots j_s k_s} = R^a_{0i_1 j_2 k_2 \cdots j_s k_s} + \partial_{i_1} \mu^a_{j_2 k_2 \cdots j_s k_s}
\ee
for some arbitrary $\mu^a_{j_2 k_2 \cdots j_s k_s}$ with Young symmetry type
$$
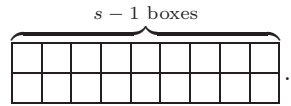
\overbrace{\yng(9,9)}^{\text{$s-1$ boxes}}.
$$
Now, the cyclic identity fulfilled by the curvature implies $\partial_{[i_1} \mu^a_{j_2 k_2] \cdots j_s k_s} = 0$, i.e., in index-free notation, $d_{(s-1)} \mu^a =0$, and this yields $\mu^a = d_{(s-1)}^{s-1} \nu^a$ for some symmetric $\nu^a_{j_2 \cdots j_s}$ \cite{DuboisViolette:1999rd,DuboisViolette:2001jk}.  Comparing with (\ref{ambiguity}), we see that this is just the ambiguity in $R^a_{0i_1 j_2 k_2 \cdots j_s k_s}$ due to the presence of $h^a_{0j_2 \cdots j_s}$.  Therefore, one can absorb $\mu^a_{j_2 k_2 \cdots j_s k_s}$ in a redefinition of the pure gauge variables $h^a_{0j_2 \cdots j_s}$ and get thereby the equations (\ref{key4}).

It is in the form (\ref{key5}) that we shall derive the twisted self-duality conditions from a variational principle.

\section{Variational principle}
\label{VarPri}
\setcounter{equation}{0}
\subsection{Prepotentials}
The searched-for variational principle involves as basic dynamical variables not the fields $h^a_{i_1 \cdots i_s}$, which are constrained, but rather ``prepotentials" that solve the constraints (\ref{constraints0}) and can be varied freely in the action.  The general solution of the constraint equation $\bar{G}^{a \, i_1 \cdots i_{s-2}} = 0$ was worked out in \cite{Henneaux:2015cda} and implies the existence of prepotentials $Z^a_{i_1 \cdots i_s}$ from which $h^a_{i_1 \cdots i_s}$ derives, such that the Einstein tensor $G^{a\,  i_1 \cdots i_s}$ of $h^a_{i_1 \cdots i_s}$ is equal to the Cotton tensor $D^{a\, i_1 \cdots i_s}$ of $Z^a_{i_1 \cdots i_s}$.

The Cotton tensor $D^{a\,  i_1 \cdots i_s}$ involves $2s-1$ derivatives of the prepotentials and possesses the property of being invariant under spin-$s$ diffeomorphisms and Weyl symmetries,
\be
\delta Z^a_{i_1 \cdots i_s} = s \partial_{(i_1} \rho^a_{i_2 \cdots i_s)} + \frac{s(s-1)}{2} \delta_{(i_1 i_2} \sigma^a_{i_3 \cdots i_s)}. \label{gaugeZ}
\ee
It is symmetric, transverse and traceless.  It was introduced for general spins in \cite{Damour:1987vm,Pope:1989vj} and \cite{Henneaux:2015cda} (where it was denoted $B$), and used extensively in three-dimensional higher spin models in \cite{Bergshoeff:2009tb,Bergshoeff:2011pm,Nilsson:2013tva,Nilsson:2015pua,Linander:2016brv,Kuzenko:2016bnv,Kuzenko:2016qdw}.  

Explicitly, the Cotton tensor of $Z^a_{i_1 \cdots i_s}$ is given by
\begin{eqnarray}
D^{ai_1 i_2  \cdots i_s} [Z] &=& \varepsilon^{i_1 j_1 k_1}\varepsilon^{i_2 j_2 k_2} \cdots \varepsilon^{i_{s-1} j_{s-1} k_{s-1}} \nonumber \\
&& \hspace{.5cm} \partial_{j_1} \partial_{j_2} \cdots \partial_{j_{s-1}} S^{a\hspace{1.38cm} i_s}_{k_1 k_2 \cdots k_{s-1}} [Z] \nonumber
\end{eqnarray}
where $S^{a \, i_1 \cdots i_s}[Z] $ is the Schouten tensor of $Z^a_{i_1 \cdots i_s}$, related to the Einstein tensor of $Z^a_{i_1 \cdots i_s}$ through
\begin{eqnarray}
S^{a \, i_1 \cdots i_s}[Z] &=& G^{a \, i_1 \cdots i_s} [Z] \nonumber \\
&& + \sum_{n=1}^{[\frac{n}{2}]} c_n  \delta^{(i_1 i_2}  \cdots \delta^{i_{2n-1} i_{2n}} G_{[n]}^{a \, i_{2n+1}  \cdots i_s)} [Z] \hspace{.5cm} \nonumber
\end{eqnarray}
with 
$$
c_n = \frac{(-1)^n}{4^n} \frac{s}{n!} \frac{(s-n-1)!}{(s-2n)!}, \; \; \; (n\geq 1)
$$
(see \cite{Henneaux:2015cda}).

Because of the gauge symmetries, the solution of the equation $G^{a\,  i_1 \cdots i_s}[h]= D^{a\,  i_1 \cdots i_s}[Z]$ for $h^a_{i_1 \cdots i_s}$ involves ambiguities.  To any given solution $h^a[Z]$ one can add an arbitrary variation of $h^a_{i_1 \cdots i_s}$ under spin-$s$ diffeomorphisms.   Furthermore $Z^a_{i_1 \cdots i_s}$ and $Z^a_{i_1 \cdots i_s} + \delta Z^a_{i_1 \cdots i_s}$ (with $\delta Z^a_{i_1 \cdots i_s}$ given by (\ref{gaugeZ})) yield $h^a[Z]$'s that differ by a spin-$s$ diffeomorphism. 

The expression for the spin-$s$ field $h_{i_1 \cdots i_s}$ in terms of the prepotential $Z_{i_1 \cdots i_s}$ contains $s-1$ derivatives in order to match the number of derivatives ($s$) of the Einstein tensor $G[h]$  with the number of derivatives ($2s-1$) of the Cotton tensor $D[Z]$. This number is odd (even) when $s$ is even (odd) and therefore, in order to match the indices of $h_{i_1 \cdots i_s}$ with those of $\partial_{k_1} \cdots \partial_{k_{s-1}} Z_{j_1 \cdots j_s}$, one needs one $\epsilon^{ijk}$ when $s$ is even and  no $\epsilon^{ijk}$ when it is odd, together with products of $\delta_{ij}$'s. 

\subsubsection{Even Spins}

We first turn to the even $s$ case.  We recall that in the spin-$2$ case, a particular solution is given by \cite{Henneaux:2004jw}
\begin{eqnarray}
h_{ij} &=& \epsilon_{(i\vert kl} \partial^k Z^l_{\ \vert j)} .
\end{eqnarray}
where the indices between the symbol $\vert \; \; \vert$ are omitted in the symmetrization -- which is as usual carried with weight one such that it is a projector.  The gauge freedom of the prepotential is given by 
\begin{eqnarray}
\delta Z_{ij} &=&
\delta_{ij} \sigma 
\ + \ 2 \ \partial_{(i} \rho_{j)} ,
\end{eqnarray}
\noindent which generates the particular diffeomorphism $\delta h_{ij} = \partial_{(i} \theta_{j)}$ of the field, where $\theta_i = \epsilon_{ikl} \partial^k \rho^l$ (it is a diffeomorphism whose parameter is divergenceless).  The generalization of this formula to general even spin $s=2n$ is given in Appendix \ref{EvenSpin}.

We give here for definiteness the expression of the spin $4$ field $h_{ijkl}$ in terms of its prepotential $\phi_{ijkl}$.  One has
\begin{eqnarray}
h_{ijkl} &=&
\epsilon_{(i\vert mn} \partial^m \left[
- \ \Delta Z^n_{\phantom{n} \vert jkl)}
\ + \  \frac{1}{2} \ \delta_{\vert jk} \Delta \bar{Z}^n_{\phantom{n}l)}
\right. \nonumber \\
&& \left. - \ \frac{1}{2} \ \delta_{\vert jk} \partial^p \partial^q Z^n_{\phantom{n} l)pq} 
\right] . \label{hZ4}
\end{eqnarray}
The gauge freedom of the prepotential is given by :
\begin{eqnarray}
\delta Z_{ijkl} &=& 
4 \ \partial_{(i} \rho_{jkl)}
\ + \ 6 \ \delta_{(ij} \sigma_{kl)} ,
\end{eqnarray}
which implies :
\be
\delta h_{ijkl} =
\partial_{(i} \theta_{jkl)} ,
\ee
where 
\begin{eqnarray}
\theta_{ijk} &=&
\epsilon_{(i\vert mn} \partial^m \mu^n_{\phantom{n} \vert jk)} ,
\\
\mu_{ijk} &=&
- \ 3 \  \Delta \rho_{ijk}
\   + \ \frac{1}{2}\delta_{(ij} \left[  
\Delta \bar{\rho}_{k)}
\ - \ \partial^p \partial^q \rho_{k)pq} 
\ \right. \nonumber \\
&& \left. \hspace{2cm} - \ 4 \  \partial^p  \sigma_{k) p} \right] .
\end{eqnarray}
In fact, as discussed in Appendix \ref{EvenSpin}, the expression (\ref{hZ4}) is, up to a multiplicative factor,  the only one (with the requested number of derivatives) that implies that a gauge variation of $Z$ gives a gauge variation of $h$.

\subsubsection{Odd Spins}
In the odd spin case, the number of derivatives on the prepotential is even, so that the expression relating $h$ to $Z$ does not involve the Levi-Civita tensor.  The expression for the spin-$3$ field in terms of its prepotential is explicitly given in \cite{Henneaux:2015cda}
\begin{eqnarray}
h_{ijk} &=& - \Delta Z_{ijk} + \frac34 \delta_{(ij} \Delta \bar{Z}_{k)} \nonumber \\
&& - \frac34 \delta_{(ij} \partial^r \partial^s Z_{k)rs} + \frac{3}{10} \delta_{(ij} \partial_{k)} \partial^r \bar{Z}_r . \label{SolG=B}
\end{eqnarray}
The last term in (\ref{SolG=B}) is actually not necessary but included so that $\delta h_{ijk} = 0$ under Weyl transformations of $Z$.   One easily verifies that a gauge transformation of the prepotential induces a gauge transformation of the spin-$3$ field.

The expression of $h_{i_1 \cdots i_s}$ in terms of $Z_{i_1 \cdots i_s}$ is given in  Appendix \ref{OddSpin} for general odd spin.

\subsection{Twisted self-duality and prepotentials}
In terms of the prepotentials, the electric fields are given by
\be
{\mathcal E}^{a \, i_1 \cdots i_s} = D^{a i_1 \cdots i_s}[Z]
\ee
while the magnetic fields have the property 
\be
\epsilon_j^{\; \;  i_1k} \partial_{k}{\mathcal B}^{a \, ji_2 \cdots i_s} = \dot{D}^{a i_1 \cdots i_s}[Z]
\ee
It follows that the twisted self-duality conditions take the form
\be
\epsilon^{i_1}_{\; \; jk} \partial^{j} D^{a \, k i_2 \cdots i_s}[Z] = \epsilon^{a}_{\; \; b} \dot{D}^{b i_1 \cdots i_s}[Z]  \label{key6}
\ee in terms of the prepotentials: the curl of the Cotton tensor of one prepotential is equal to ($\pm$) the time derivative of the other.  

\subsection{Action}
In their form (\ref{key6}), the twisted self-duality conditions are easily checked to derive from the following variational principle,
\be
S[Z] = \int dx^0 \left[ \int d^3x \ \frac{1}{2}\varepsilon_{ab} D^{a\, i_1 \cdots i_s} \dot{Z}^b_{i_1 \cdots i_s} - H \right] \label{ActionPrepot00}
\ee
where the Hamitonian $H$ reads
\be
H = \int d^3x \delta_{ab} \left(\sum_{k=0}^{[\frac{s}{2}]} a_k  G^{[k] a \, i_1 \cdots i_{s-2k}}G^{[k] b}_{\; \; \; \; \; \; \; i_1 \cdots i_{s-2k}}  \right) \label{HAM11}
\ee
where $G^{[k] a \, i_1 \cdots i_{s-2n}}$ stands for the $k$-th trace of the Einstein tensor  $G^{a \, i_1 \cdots i_s}[Z]$ of the prepotential $Z^a_{i_1 \cdots i_s}$. A lengthy but conceptually direct computation shows that the coefficients  $a_k$ are explicitly given by
$$
 a_k =  \left( - \right)^k \frac{n!}{\left(n-k\right)!k!}
\frac{\left(2n-k-1\right)! \left(2n-1\right)!! }{2^k \left(2n-1\right)!\left(2n-2k-1\right)!!} \frac{1}{2}
$$
for even spin $s=2n$, and 
$$
a_k =
\left( - \right)^k \frac{n!}{\left(n-k\right)!k!}
\frac{\left(2n-k\right)! \left(2n+1\right)!! }{2^k \left(2n\right)!\left(2n-2k+1\right)!!} \frac{1}{2}
$$ for odd spin $s = 2n +1$, where the definition of the double factorial is recalled in the appendix.
These coefficients are in fact uniquely determined up to an overall factor by the property that the action is invariant, up to a surface term, under the gauge symmetries (\ref{gaugeZ}) of the prepotentials.  Invariance under spin-$s$ diffeomorphisms is manifest, while invariance under spin-$s$ Weyl symmetry forces $a_k$ to be given by the above expression (up to an overall factor).

We close this section by observing that the Hamiltonian (\ref{HAM11}) can be rewritten more simply as:
\be
H =  \frac{1}{2} \int d^3x \, \delta_{ab} D^{a\, i_1 \cdots i_s} \, {\epsilon^{mn}}_{i_1}\, \partial_m{Z}^b_{n i_2 \cdots i_s},
\ee
an expression that more clearly exhibits its gauge invariance\footnote{ We are grateful to Victor Lekeu for pointing this out to us.}.

\section{Hamiltonian formalism}
\label{Hamiltonian}
\setcounter{equation}{0}

\subsection{Constraints and Hamiltonian}
It turns out that the action (\ref{ActionPrepot00}) is exactly the action that one obtains by rewriting the Fronsdal action action in Hamiltonian form and solving the constraints. 

The procedure to establish this fact proceeds as follows.  
\begin{enumerate}
\item First one writes the Fronsdal action in Hamiltonian form \cite{Metsaev:2011iz,Campoleoni:2016uwr}.  The Hamiltonian canonical variables are the spatial components $h_{i_1 \cdots i_s}$ of the spin-$s$ field, their conjugate momenta $\pi^{i_1 \cdots i_s}$, the variables $\alpha_{i_1 \cdots i_{s-3}}$ equal to $h_{000 i_1 \cdots i_{s-3}} - 3 \delta^{kl} h_{0 i_1 \cdots i_{s-3} kl}$ and their conjugate momenta $\tilde{\Pi}^{i_1 \cdots i_{s-3}}$.  The canonical action takes the form
\begin{eqnarray}
&& S[h_{i_1 \cdots i_s}, \pi^{i_1 \cdots i_s}, \alpha_{i_1 \cdots i_{s-3}}, \tilde{\Pi}^{i_1 \cdots i_{s-3}}, {\mathcal N}^{i_1 \cdots i_{s-2} } , N^{i_1 \cdots i_{s-1} }] \nonumber \\
&& = \int dx^0\left[ \int d^3x (\pi^{i_1 \cdots i_s} \dot{h}_{i_1 \cdots i_s} + \tilde{\Pi}^{i_1 \cdots i_{s-3}} \dot{\alpha}_{i_1 \cdots i_{s-3}})  \right. \nonumber \\
&& \left. - H   
  - \int d^3 x ({\mathcal N}^{i_1 \cdots i_{s-2} } {\mathcal C}_{i_1 \cdots i_{s-2}}  \right. \nonumber \\ 
  && \left. \hspace{3cm} +N^{i_1 \cdots i_{s-1} } {\mathcal C}_{i_1 \cdots i_{s-1}}  ) \right]
\end{eqnarray}
where ${\mathcal C}_{i_1 \cdots i_{s-2}}$ and ${\mathcal C}_{i_1 \cdots i_{s-1}}$ are the constraint-generators related to temporal ($\epsilon_{0 i_1 \cdots i_{s-2}}$) and spacelike ($\epsilon_{ i_1 \cdots i_{s-1}}$) spin-$s$ diffeomorphisms, respectively, and ${\mathcal N}_{i_1 \cdots i_{s-2} } = h_{00i_1 \cdots i_{s-2}}$ and $N_{i_1 \cdots i_{s-1} } = h_{0i_1 \cdots i_{s-1}}$ are the corresponding Lagrange multipliers. The explicit form of the constraints is rather cumbersome and has been given in  \cite{Campoleoni:2016uwr}.  The function $H$ is the Hamiltonian.  It is the integral over space of a density ${\mathcal H}$ which is quadratic in the conjugate momenta and in the first spatial derivatives of $h_{i_1 \cdots i_s}, \alpha_{i_1 \cdots i_{s-3}}$, $H = \int d^3x {\mathcal H}$. We shall not need here the explicit expression of ${\mathcal H}$ in terms of $\partial_k h_{i_1 \cdots i_s}, \pi^{i_1 \cdots i_s}, \partial_k \alpha_{i_1 \cdots i_{s-3}}, \tilde{\Pi}^{i_1 \cdots i_{s-3}}$, which is also cumbersome.
By anology with the spin-$2$ case, we shall call the constraint
\be
{\mathcal C}_{i_1 \cdots i_{s-2}} = 0
\ee
the ``Hamiltonian constraint" and the constraint
\be
{\mathcal C}_{i_1 \cdots i_{s-1}} = 0
\ee
the ``momentum constraint".
\item The second step is to solve the constraints in terms of prepotentials $Z^a_{i_1 \cdots i_s}$, $s=1,2$.     One needs two prepotentials, one for solving the Hamiltonian constraint, the other one for solving the momentum constraint.  The procedure uses the conformal tools developed in \cite{Henneaux:2015cda} and follows exactly the same pattern as for spins $2$ and $3$. It is displayed in the next two sections. 
\item One then inserts the expression for the canonical variables in terms of the prepotentials inside the action and obtains (\ref{ActionPrepot00}) 
\end{enumerate}

\subsection{Solving the momentum constraint}
We first solve the momentum constraint.  This constraint reads \cite{Campoleoni:2016uwr}
\be
{\mathcal C}^{i_1 \cdots i_{s-1}} \equiv \partial_k \pi^{ki_1 \cdots i_{s-1}} + \hbox{``more"}= 0
\ee
where ``more" stands for terms that are linear in the second order derivatives of $\alpha_{i_1 \cdots i_{s-3}}$, which one can set to zero by a suitable gauge transformation.  In the gauges where ``more" vanishes, the constraint reduces to 
\be
\partial_k \pi^{ki_1 \cdots i_{s-1}} = 0,
\ee
the general solution of which is given by $\pi^{i_1 \cdots i_{s}} = G^{i_1 \cdots i_{s}}[P]$ \cite{DuboisViolette:1999rd,DuboisViolette:2001jk}.  Here $G^{i_1 \cdots i_{s}}[P]$ is the Einstein tensor of some prepotential  $P_{i_1 \cdots i_s}$ which is totally symmetric.

For fixed momentum $\pi^{i_1 \cdots i_s}$, the prepotential $P_{i_1 \cdots i_s}$ is determined up to a spin-$s$ diffeomorphism.  However, there is a residual gauge freedom in the above gauges, so that $\pi^{i_1 \cdots i_s}$ is not completely fixed.  It is straightforward but somewhat tedious to check that the residual gauge symmetry is accounted for by a spin-$s$ Weyl transformation of the prepotential $P_{i_1 \cdots i_s}$, which therefore enjoys all the gauge symmetries of a conformal spin-$s$ field.

These results extend what was  found earlier  for spins $2$ \cite{Henneaux:2004jw} and $3$ \cite{Henneaux:2015cda}.  It is instructive to exhibit explicitly the formulas in the case of spin $4$,  which illustrates all the points and still yields readable expressions.

The momentum constraint reads in this case
\begin{eqnarray} 
0 &=& 
4 \ \partial^n \pi_{klmn}
\ + \ 6 \ \delta_{(kl} \Delta \alpha_{m)} \nonumber \\
& - & 10 \ \delta_{(kl} \partial_{m)} \partial^n \alpha_n . \label{constr_2}
\end{eqnarray}
and the  gauge freedom is:
\begin{eqnarray}
\delta \pi_{klmn} &=&
- \ 12 \ \partial_{(k} \partial_l \Xi_{mn)} \nonumber \\
&& + \ 12 \ \delta_{(kl} \left( \Delta \Xi_{mn)}\ + \ \partial_m \partial^p \Xi_{n)p} \right)
\nonumber \\ 
& + & 4 \ \delta_{(kl} \delta_{mn)} \left(2 \ \Delta \bar{\Xi} \ + \ 3 \ \partial^p \partial^q \Xi_{pq} \right) ,
\\
\delta \alpha_k &=& 
- \ 6 \ \partial^l \Xi_{kl}
\ - \ 2 \ \partial_k \bar{\Xi} .
\end{eqnarray}

The residual gauge transformations in the gauge $3 \ \Delta \alpha_k \ - \ 5 \ \partial_k \partial^l \alpha_l = 0$, which eliminates the $\alpha$-terms from the constraint,  must fulfill
$$
0 = - \ 18 \ \partial^l \Delta \Xi_{kl}
\ + \ 4 \ \partial_k \Delta \bar{\Xi}
\ + \ 30 \ \partial_k \partial^l \partial^m \Xi_{lm} .
$$
The divergence of this equation gives (after acting with $\Delta^{-1}$), $
3 \ \partial^k \partial^l \Xi_{kl}
\ + \ \Delta \bar{\Xi} = 0$.
Substituting this finding in the previous equation yields, after acting again with $\Delta^{-1}$, 
$3 \ \partial^l \left(\Xi_{kl} \ + \ \frac{1}{3} \ \delta_{kl} \bar{\Xi}\right) = 0$.
This is the divergence of a symmetric tensor, so the solution is the double divergence of a tensor with the symmetry of the Riemann tensor :
\begin{eqnarray}
\Xi_{kl} \ + \ \frac{1}{3} \ \delta_{kl} \bar{\Xi} &=&
\partial^m \partial^n \Theta_{mknl} .
\end{eqnarray}
Therefore, one has
\begin{eqnarray}
\Xi_{kl} &=&
\partial^m \partial^n \Theta_{mknl}
\ - \ \frac{1}{6} \ \delta_{kl} \partial^m \partial_n \Theta_{mp}^{\phantom{mp}np}.
\end{eqnarray}
This class of gauge transformations can be checked  to give a zero variation not only to the contribution of $\alpha_k$ to the constraint but in fact also to $\alpha_k$ itself.

We can dualize $\Theta_{klmn} = \epsilon_{klp}\epsilon_{mnq} \theta^{pq}$, with a symmetric $\theta_{kl}$, to obtain :
\begin{eqnarray}
\Xi_{kl} &=&
\frac{5}{6} \ \delta_{kl} \left(\Delta \bar{\theta} 
\ - \ \partial^m \partial^n \theta_{mn}\right) \nonumber \\
& + &\ 2 \ \partial_{(k} \partial^m \theta_{l)m}
\ - \ \partial_k \partial_l \bar{\theta}
\ - \ \Delta \theta_{kl} . \qquad
\end{eqnarray}
The gauge transformation of $\pi_{klmn}$ with this parameter is found then to be exactly the Einstein tensor of a Weyl diffeomorphism
\begin{eqnarray}
\delta \pi_{klmn} &=&
G_{klmn} \left[12 \ \delta_{(pq} \theta_{rs)}\right] .
\end{eqnarray}

Once the spin-$4$ momentum constraint has been brought in the standard form $\partial^k \pi_{klmn} = 0$ by partial gauge fixing, it can be solved by the familiar techniques recalled at the beginning of this section for general $s$.  As it is known \cite{DuboisViolette:1999rd,DuboisViolette:2001jk}, the general solution of the equation $\partial^k \pi_{klmn} = 0$ is indeed the Einstein tensor of a symmetric tensor $P_{klmn}$, which is the prepotential for the momenta: 
$$\pi_{klmn} = G_{klmn} \left[ P \right].$$ 
Since the gauge freedom of $\pi_{klmn}$ is given by the Einstein tensor of a Weyl transformation, the gauge freedom of the prepotential $P_{klmn}$ will be given by a spin-$4$ Weyl transformation and a spin-$4$ diffeomorphism:
\begin{eqnarray}
\delta P_{klmn} &=&
4 \ \partial_{(k} \xi_{lmn)}
\ + \ 6 \ \delta_{(kl} \lambda_{mn)} ,
\end{eqnarray}
\noindent the first term corresponding to all the transformations of the prepotential leaving $\pi_{klmn}$ invariant and the second to those realizing on it a residual gauge transformation.

The solution in a general gauge will be given by these expressions to which are added general gauge transformation terms, parametrized by further prepotentials that drop from the action by gauge invariance and which are usually not considered for that reason.

\subsection{Solving the Hamiltonian constraint}
We now solve the Hamiltonian constraint.  The functions ${\mathcal C}_{i_1 \cdots i_{s-2}}$ are linear in the second order derivatives of the spin-$s$ field $h_{i_1 \cdots i_s}$ and linear in the first order derivatives of the conjugate momenta $\tilde{\Pi}^{i_1 \cdots i_{s-3}}$.   One may rewrite these constraints in the equivalent form (suppressing indices) 
\be
\Psi - d_{(s-2)} \tilde{\Pi} = 0 \label{HamBis}
\ee
where $\Psi$ is the function of the second order derivatives of the spin-$s$ field with Young symmetry
$$\Psi <> \overbrace{\yng(8)}^{(s-2) \text{ boxes}},$$
introduced in \cite{Henneaux:2015cda}, such that $\! ^* \bar{G}  \equiv  \! ^*  G^{[1]}= d_{(s-2)}^{s-2} \Psi $, where $^*$ denotes here the dual on {\em all} indices.  

The results of \cite{DuboisViolette:1999rd,DuboisViolette:2001jk} applied to $d_{(s-2)}$ such that $d_{(s-2)}^{s-1}=0$ imply that the equation (\ref{HamBis}) is completely equivalent to $ d_{(s-2)}^{s-2} \Psi = 0$, i.e., $G^{[1]}[h] = 0$.  As we have recalled, this equation implies in turn  the existence of a prepotential $Z$ for $h$ (continuing to omit indices) such that the Einstein tensor of $h$ is equal to the Cotton tensor of $Z$. Once $h$ is expressed in terms of $Z$, the expression $ d_{(s-2)}^{s-2} \Psi[Z]$ identically vanishes, implying according to  the generalized algebraic Poincar\'e lemma of \cite{DuboisViolette:1999rd,DuboisViolette:2001jk} that one can write $\Psi[Z] = d_{(s-2)} \tilde{\Pi}[Z]$, for some $\tilde{\Pi}[Z]$.  

This completely solves the Hamiltonian constraint in terms of the prepotential $Z_{i_1 \cdots i_2}$.   By construction, this prepotential has the gauge symmetry of a conformal spin-$s$ field.

The procedure is direct for spins $1$, where there is no Hamiltonian constraint,  and $2$, where one gets directly $G^{[1]}[h] = 0$ without having to differentiate. It was detailed for spin $3$ in \cite{Henneaux:2015cda}.   It is again instructive to illustrate it explicitly for the spin $4$ field, where the formulas remain readable. 

The hamiltonian constraint for the spin $4$ is :
\begin{eqnarray}
\mathcal{C}_{kl} &\equiv& 
3 \ \partial_{(k} \tilde{\Pi}_{l)}
\ + \ \delta_{kl} \partial^m \tilde{\Pi}_m \nonumber \\
& - & 6 \ \left(
\Delta \bar{h}_{kl}
\ - \ \partial^m \partial^n h_{klmn}
\ + \ \partial_{(k} \partial^m \bar{h}_{l)m}
\ + \ \partial_k \partial_l \bar{\bar{h}} \right) \nonumber \\
& - & 4 \ \delta_{kl} \Delta \bar{\bar{h}}  = 0 , \qquad \label{constr_1}
\end{eqnarray}
The gauge freedom of $ h_{klmn}$ and $\tilde{\Pi}_k$  is :
\begin{eqnarray}
\delta \tilde{\Pi}_k &=& 
6 \ \Delta \bar{\xi}_k \ + \ 10 \ \partial_k \partial^l \bar{\xi}_l ,
\\
\delta h_{klmn} &=& 
4 \ \partial_{(k} \xi_{lmn)} .
\end{eqnarray}
One can equivalently rewrite the constraint as
\be
\mathcal{C}_{kl} \ - \ \frac{1}{6} \ \delta_{kl} \bar{\mathcal{C}} \ \equiv \ 3 \ \partial_{(k} \tilde{\Pi}_{l)} \ - \ 6 \ \Psi_{kl} = 0,
\ee
where $\bar{\mathcal{C}}$ is the trace of $\mathcal{C}_{kl}$ and
\begin{eqnarray}
\Psi_{kl} &\equiv&
\Delta \bar{h}_{kl}
\ - \ \partial^m \partial^n h_{klmn} \nonumber \\
&+ & \partial_{(k} \partial^m \bar{h}_{l)m}
\ + \ \partial_k \partial_l \bar{\bar{h}} .
\end{eqnarray}
One has 
\begin{eqnarray}
\epsilon_{k mp} \epsilon_{l nq} \partial^m \partial^n \Psi^{pq} &=&
\bar{G}_{kl} \left[ h \right]. \qquad \qquad
\end{eqnarray}
where $\bar{G}_{kl}$ is the trace of the Einstein tensor $G_{klmn} \left[ h \right]$ of $h_{klmn}$. So, one gets:
\begin{eqnarray}
\bar{G}_{kl} \left[h\right] &=&
\ - \ \frac{1}{6} \ \epsilon_{k mp} \epsilon_{l nq} \partial^m \partial^n \tilde{\mathcal{C}}^{pq}  .
\end{eqnarray}
with $\tilde {\mathcal C}^{pq} \equiv \mathcal{C}^{pq} \ - \ \frac{1}{6} \ \delta^{pq} \bar{\mathcal{C}}$.

The Hamiltonian constraint implies $\bar{G}_{kl} \left[ h \right] = 0$.  Therefore, the theorem of \cite{Henneaux:2015cda} yields $G_{ijkl} \left[h\right] = B_{ijkl} \left[Z\right]$  for some prepotential $Z_{klmn}$, where $B_{ijkl}$ is the Cotton tensor.  Explicitly,
\begin{eqnarray}
h_{ijkl}\left[Z\right] &=&
\epsilon_{(i\vert mn} \partial^m \left[
- \ \Delta Z^n_{\phantom{n} \vert jkl)}
\ + \  \frac{1}{2} \ \delta_{\vert jk} \Delta \bar{Z}^n_{\phantom{n}l)} \right. \nonumber \\
 & - &\left.  \frac{1}{2} \ \delta_{\vert jk} \partial^p \partial^q Z^n_{\phantom{n} l)pq} 
\right] + \ 4 \ \partial_{(i} \omega_{jkl)}, \qquad
\end{eqnarray}
where we have added a spin-$4$ diffeomorphism term parametrized by $\omega_{jkl}$.

Direct substitution gives then
\begin{eqnarray}
\Psi_{ij}  &=&
\partial_{(i\vert} \left[
3 \  \Delta \bar{\omega}_{j)}
\ + \ 5 \ \partial_{\vert j)} \partial^k \bar{\omega}_k \right. \nonumber \\
& - & \left. \frac{1}{8} \ \epsilon_{\vert j ) mn} \partial^m \left(
 \partial^k \Delta \bar{Z}^n_{\phantom{n}k}
\ + \ \partial^p \partial^q \partial^k Z^n_{\phantom{n} kpq} \right)
\right] . \qquad \qquad
\end{eqnarray}
from which one derives, using the constraint, the following expression for $\tilde{\Pi}_i$,
\begin{eqnarray}
\tilde{\Pi}_i &=&
6 \  \Delta \bar{\omega}_{i}
\ + \ 10 \ \partial_{i} \partial^k \bar{\omega}_k \nonumber \\
& - & \frac{1}{4} \ \epsilon_{i mn} \partial^m \left(
  \partial^k \Delta \bar{Z}^n_{\phantom{n}k}
\ + \ \partial^p \partial^q \partial^k Z^n_{\phantom{n} kpq} \right) 
, \qquad \qquad
\end{eqnarray}
One could in fact add  a solution $\kappa_i$ of the equation $\partial_{(i} \kappa_{j)} = 0$ (Killing equation) to $\tilde{\Pi}_i $ but we do not consider this possibility here by assuming for instance appropriate boundary conditions (vanishing of all fields at infinity eliminates $\kappa_i \sim C_i + \mu_{ij} x^j$, where $C_i$ and $\mu_{ij} = - \mu_{ji}$ are constants).

\subsection{Hamiltonian Action in terms of prepotentials}

The elimination of the canonical variables in terms of the prepotentials in the action is a rather burdensome task.  However, the derivation can be considerably short-cut  by invariance arguments.

The kinetic term in the action is quadratic in the prepotentials $Z$ and $P$ and involves $2s -1$ spatial derivatives, and one time derivative.  Furthermore, it must be invariant under spin-$s$ diffeomorphisms and spin-$s$ Weyl transformations of both prepotentials.  This implies, making integrations by parts if necessary, that the kinetic term has necessarily the form of the kinetic term of the action (\ref{ActionPrepot00}) upon identification of the prepotential $Z$ with the prepotential $Z^1$ and the prepotential $P$ with the prepotential $Z^2$. 

Similarly, the Hamiltonian is the integral of a quadratic expression in the prepotentials $Z$ and $P$ involving $2s$ spatial derivatives.  By spin-$s$ diffeomorphism invariance,  it can be written as the integral of a quadratic expression in their Einstein tensors and their successive traces  -- or equivalently, the Schouten tensors and their successive traces.   As we have seen, Weyl invariance fixes then the coefficients up to an overall rescaling, so that the Hamiltonian takes necessarily the form (\ref{HAM11}), but with $\delta_{ab}$ that might be replaced by a diagonal $\mu_{ab}$ with eigenvalues different from $1$.  However, given that the equations of motion following from  the action (\ref{ActionPrepot00})  are, as we have seen, consequences of the Fronsdal equations, they cannot be in contradiction with  the equations of motion following from the Hamiltonian action (equivalent to the Fronsdal equations) and this is possible only if $\mu_{ab} = \delta_{ab}$. 

One thus concludes that the action (\ref{ActionPrepot00}) for twisted self-duality is indeed the Hamiltonian form of the Fronsdal action with the constraints solved for in terms of prepotentials.  This shows in particular that the electric field-magnetic field version of twisted self-duality obtained by considering only the spatial components of (\ref{key1}) form indeed a complete set, as announced  The analysis also shows that duality-conjugate and canonically conjugate are equivalent (up to field-independent factors)

\section{Additional considerations}
\setcounter{equation}{0}
\label{Additional}

\subsection{Higher dimensions and twisted self-duality}
In higher spacetime dimension $D$,  the equations of motion can also be reformulated as twisted self-duality conditions on the curvatures of the spin-$s$ field and its dual. What is new is that the dual of a spin-$s$ field is not given by a symmetric tensor, but by a tensor of mixed Young symmetry type
$$
^\text{$D-3$ boxes}  \overbrace{\left\{ \yng(10,1,1,1,1,1,1) \right.}^{\text{$s$ boxes}}. 
$$
Consequently, the curvature tensor and its dual are also tensors of different types.  Nevertherless, the electric (respectively, magnetic) field of the spin-$s$ field is a spatial tensor of the same type as the magnetic (respectively, electric) field of its dual and  the twisted self-duality conditions again equate them (up to $\pm$ similarly to Eq. (\ref{key3})).  
The electric and magnetic fields are subject to tracelessness constraints that can be solved in terms of appropriate prepotentials, which are the variables for the variational principle from which the twisted self-duality conditions derive.  Again, this variational principle is equivalent to the Hamiltonian variational principle.

We have not worked out the specific derivation for all spins in higher dimensions $D$, but the  results of \cite{Bunster:2013oaa} for the spin-$2$ case, together with our above analysis, make us confident that this derivation indeed goes through as described here.

\subsection{Manifest $SO(2)$ electric-magnetic duality invariance in $D=4$}
In $D=4$ spacetime dimensions, the spin-$s$ field and its dual are tensors of the same type, as we have seen.  The equations enjoy then $SO(2)$ electric-magnetic duality invariance that rotates the field and its dual in the internal two-dimensional space that they span.  This comes over and above the twisted self-duality reformulation. 

The $SO(2)$ electric-magnetic duality invariance amounts to perform rotations in the internal space of the prepotentials.   It is clear that it is also a symmetry of the action (\ref{ActionPrepot00}).  Thus, the prepotential reformulation makes it obvious that $SO(2)$ electric-magnetic duality invariance is a manifest off-shell symmetry, and not just a symmetry of the equations of motion.

\section{Comments and Conclusions}
\label{Conclusions}
\setcounter{equation}{0}
 
In this paper, we have achieved two things.  (i) First, we have rewritten the equations of motion for higher spin gauge fields as twisted self-duality conditions, in which the spin-$s$ field and its dual are put on exactly the same footing.  (ii) Second,  by introducing prepotentials for the spin-$s$ field and its dual, we have shown how these equations derive from a variational principle, providing thereby a duality symmetric formulation of the theory.  

One observes again, for all spins, the intriguing emergence of higher spin Weyl gauge symmetries \cite{Fradkin:1985am,Vasiliev:2001zy,Segal:2002gd,Shaynkman:2004vu,Metsaev:2007rw} for the prepotentials, in addition to spin-$s$ diffeomorphisms.  This generalizes what was found in the spin-$2$ case in \cite{Henneaux:2004jw}.

One should also stress the remarkable simplicity of the final action.  Furthermore, this final action  takes the same form for all spins. This uniformity suggests that the prepotential formalism might perhaps be a good starting point for exploring the symmetries mixing all spins  -- in particular the $sp(8)$-symmetry  in four spacetime dimensions \cite{Fronsdal:1985pd,Gelfond:2015poa}.

We have restricted the analysis to massless fields in flat spacetime.  Extension to constant curvature backgrounds \cite{Julia:2005ze} and to partially massless fields would be of definite interest \cite{Deser:2012qg,Deser:2012qg,Hinterbichler:2016fgl}.

\section*{Acknowledgments} 
 We thank Andrea Campoleoni for useful discussions. A.L. is Research Fellow at the Belgian F.R.S.-FNRS. The work of S.H. has been supported by the Fondecyt grant N\textordmasculine \; 3160781. This work was partially supported by the ERC Advanced Grants ``SyDuGraM" and ``High-Spin-Grav", by F.R.S.-FNRS through the conventions PDR-T.1025.14 and IISN-4.4503.15, and by the ``Communaut\'e Fran\c{c}aise de Belgique" through the ARC program. The Centro de Estudios Cient\'ificos (CECs) is funded by the Chilean Government through the Centers of Excellence Base Financing Program of Conicyt.

\appendix

\section{Prepotentials for even spins}
\label{EvenSpin}

We give in this appendix the form of the spin-$s$ field $h_{i_1 \cdots i_s}$ in terms of the corresponding prepotential $Z_{i_1 \cdots i_s}$ when $s$ is even.  The case of an odd $s$ is treated in Appendix \ref{OddSpin}. 

Because of the gauge symmetries, the expression $h[Z]$ is not unique.  To any solution, one may add a gauge transformation term.  Our particular solution corresponds to a definite choice.

Our strategy is as follows: (i) First, one writes the most general form for $h$ in terms of $Z$ compatible with the index structure and the fact that it contains $s-1$ derivatives. (ii) Second, one fixes the coefficients of the various terms such that a gauge transformation of $Z$ induces a gauge transformation of $h$.  This turns out to completely fix $h[Z]$ up to an overall multiplicative constant.  (iii) Third, one fixes that multiplicative constant through the condition $G[h[Z]] = D[Z]$, which we  impose and verify in a convenient gauge for $Z$.

\subsection{First Step}

A generic term in the expression for $h_{i_1 \cdots i_s}$  in terms of $Z_{i_1 \cdots i_s}$ involves one Levi-Civita tensor when $s$ is even, as well as  $s-1$ derivatives of $Z$.  It can also contain a product of $p$  $\delta_{i_j i_k}$'s with free indices among $i_1, i_2, \cdots, i_s$. Hence a generic term takes the form
\be
 \delta_{i_1 i_2} \cdots \delta_{i_{2p-1} i_{2p}}\, \epsilon_{k_1 k_2 k_3} \, \partial_{m_1} \cdots \partial_{m_{s-1}} Z_{j_1 \cdots j_s}  \label{GenericTermEven}
\ee
for some $p$ such that $0 \leq p \leq n-1$ where $s=2n$ ($p$ cannot be equal  to $n$ since the Levi-Civita symbol must necessarily carry a free index, see below, so that there must be at least one free index left). Among the indices $k_1, k_2, k_3,  m_1,  \cdots, m_{s-1}, j_1, j_2, \cdots, j_s$, there are $s-2p$ indices equal to the remaining $i_a$'s, and the other indices are contracted with $\delta^{ab}$'s.   There is also an implicit symmetrization over the free indices $i_a$, taken as before to be of weight one.

The structure of the indices of the Levi-Civita symbol is very clear: because of the symmetries, one index is a free index $i_b$, one index is contracted with a derivative operator, and one index is contracted with an index of $Z$. Furthermore, if an index $m_b$ on the derivatives is equal to one of the free indices $i_b$, then, the term can be removed by a gauge transformation.  This means that apart from one index contracted with an index of the Levi-Civita tensor, the remaining indices on the derivative operators are necessarily contracted either among themselves to produce Laplacians or with indices of $Z$. In other words, the remaining free indices, in number $s - 2p -1$ are carried by $Z$.  One index on $Z$ is contracted with one index of the Levi-Civita tensor as we have seen, and  the other indices on $Z$, in number $2p$, are contracted either among themselves to produce traces or with the indices carried by the derivative operators. Thus, if we know the number of traces that occur in $Z$, say $q$, the structure of the term (\ref{GenericTermEven}) is completely determined,
 \begin{eqnarray}
&& \hspace{-1cm} \delta_{i_1 i_2} \cdots \delta_{i_{2p-1} i_{2p}}\, \epsilon_{i_{2p+1} k} ^{\; \; \; \; \; \;  \; \; \; \; t} \, \partial^k \partial^{j_1} \cdots \partial^{j_{2p-2q}} \Delta^{n-1-p+q} \nonumber \\
&& \hspace{3cm} Z^{[q]}_{i_{2p+2}\cdots i_s t j_1 \cdots j_{2p-2q}}, \label{GenericTermEven2}
\end{eqnarray}
or, in symbolic form,
\be
\delta^p  \left(\epsilon \cdot \partial \cdot\right) \left(\partial \cdot \partial \cdot\right)^{p-q} \Delta^{n-1-p+q} Z^{[q]} 
\ee
One has $0 \leq q \leq p$ and complete symmetrisation on the free indices $i_b$ is understood.

Accordingly, the expression for $h_{i_1 \cdots i_s}$ in terms of $Z_{i_1 \cdots i_s}$ reads
\begin{eqnarray}
&& h = \nonumber \\
&& \sum_{p = 0}^{n-1}
\sum_{ q = 0}^p
a_{p,q} \ \delta^p  \left(\epsilon \cdot \partial \cdot\right) \left(\partial \cdot \partial \cdot\right)^{p-q} \Delta^{n-1-p+q} Z^{[q]} . \hspace{1cm} 
\end{eqnarray}
where the coefficients $a_{p,q}$ are determined next.

\subsection{Second Step}
By requesting that a gauge transformation of $Z_{i_1 i_2 \cdots i_s}$ induces a gauge transformation of $h_{i_1 \cdots i_s}$, the coefficients $a_{p,q}$ are found to be given by
\begin{eqnarray}
&& a_{p,q} = 2^{-p}  \left(-\right)^q  \times \nonumber \\
&&
\frac{\left(n - 1\right)!\left(2n - p - 1\right)!\left(2n- 1\right)!!}
{q!\left(p - q\right)!\left(n - p -1\right)!\left(2n - 1\right)! \left(2n - 2p -1\right)!!} a  \hspace{1cm}
\end{eqnarray}
where the multiplicative constant $a$ is undetermined at this stage. The double factorial of an odd number $2k+1$ is equal to the product of all the odd numbers up to $2k+1$,
$$
(2k+1)!! = 1 \cdot 3 \cdot 5 \cdots (2k-1) \dot (2k+1)
$$
The computation is fastidious but conceptually straightforward and left to the reader.

\subsection{Third Step}
Finally, we fix the remaining coefficient $a$ by imposing that $G[h[Z]] = D[Z]$.  This is most conveniently done in the gauge 
\be
\partial^{i_1} Z_{i_1 \cdots i_s} = 0, \; \; \; Z^{i_1}_{\; \; \; i_1 i_3 \cdots i_s} = 0
\ee
(transverse, traceless gauge). This gauge is permissible given that the gauge transformations of the prepotential involves both spin-$s$ diffeomorphisms and spin-$s$ Weyl transformations.  In that gauge, the Cotton tensor reduces to 
\begin{eqnarray}
D\left[ Z \right] _{i_1 i_2 \cdots i_s}&=&
- \ \epsilon_{(i_1 \vert jk} \, \partial^j \,  \Delta^{2n-1} Z^{k}_{\; \; \; \vert i_2 \cdots i_s)}
\end{eqnarray}
or in symbolic form,
\begin{eqnarray}
D\left[ Z \right] &=&
- \ \left(\epsilon \cdot \partial\cdot\right) \Delta^{2n-1} Z ,
\end{eqnarray}
while $h_{i_1 \cdots i_s}$ is also divergenceless and traceless (which shows, incidentally, that on the $\bar{G}[h]=0$ shell, one may impose both conditions also on $h$) and its Einstein tensor, expressed in terms of $Z$, becomes
\begin{eqnarray}
G\left[ h\left[ Z \right] \right]_{i_1 \cdots i_s} &=&
\left(-\right)^n \Delta^n h\left[ Z \right]_{i_1 \cdots i_s}
\nonumber \\ &=&
\left(- \right)^n a \  \epsilon_{(i_1 \vert jk} \, \partial^j \,  \Delta^{2n-1} Z^{k}_{\; \; \; \vert i_2 \cdots i_s)} \nonumber
\end{eqnarray}
i.e.,
\begin{eqnarray}
G\left[ h\left[ Z \right] \right] &=&
\left(- \right)^n a \ \epsilon \cdot \partial\cdot \Delta^{2n-1} Z .
\end{eqnarray}
This shows that 
\be
a = - (-)^n 
\ee
and completes the determination of $h$ in terms of its prepotential $Z$.

\section{Prepotentials for odd spins}
\label{OddSpin}

The procedure for odd spins follows the same steps, but now there is no Levi-civita tensor involved in the expression $h[Z]$ since there is an even number of derivatives.

\subsection{First Step} 

A generic term in the expression for $h_{i_1 \cdots i_s}$  in terms of $Z_{i_1 \cdots i_s}$ involves  $s-1$ derivatives of $Z$.  It can also contain a product of $p$  $\delta_{i_j i_k}$'s with free indices among $i_1, i_2, \cdots, i_s$. Hence a generic term takes the form
\be
 \delta_{i_1 i_2} \cdots \delta_{i_{2p-1} i_{2p}}\,  \partial_{m_1} \cdots \partial_{m_{s-1}} Z_{j_1 \cdots j_s}  \label{GenericTermOdd}
\ee
for some $p$ such that $0 \leq p \leq n$ where $s=2n+1$. Among the indices $m_1,  \cdots, m_{s-1}, j_1, j_2, \cdots, j_s$, there are $s-2p$ indices equal to the remaining $i_a$'s, and the other indices are contracted with $\delta^{ab}$'s.   There is also an implicit symmetrization over the free indices $i_a$.

Again, if an index $m_b$ on the derivatives is equal to one of the free indices $i_b$, then, the term can be removed by a gauge transformation.  This means that the indices on the derivative operators are necessarily contracted either among themselves to produce Laplacians or with indices of $Z$. In other words, the remaining free indices, in number $s - 2p$ are carried by $Z$.  The other indices on $Z$, in number $2p $, are contracted either among themselves to produce traces or with the indices carried by the derivative operators. Thus, if we know the number of traces that occur in $Z$, say $q$, the structure of the term (\ref{GenericTermOdd}) is completely determined, as in the even spin case
 \begin{eqnarray}
&& \hspace{-1cm} \delta_{i_1 i_2} \cdots \delta_{i_{2p-1} i_{2p}}\, \partial^{j_1} \cdots \partial^{j_{2p-2q}} \Delta^{n-p+q} \nonumber \\
&& \hspace{3cm} Z^{[q]}_{i_{2p+1}\cdots i_s  j_1 \cdots j_{2p-2q}}, \label{GenericTermOdd2}
\end{eqnarray}
or, in a more compact way :
\begin{eqnarray}
 \delta^p   \left(\partial \cdot \partial \cdot\right)^{p-q} \Delta^{n-p+q} Z^{[q]} .
\end{eqnarray}
One has $0 \leq q \leq p$ and complete symmetrisation on the free indices $i_b$ is understood.

Accordingly, the expression for $h_{i_1 \cdots i_s}$ in terms of $Z_{i_1 \cdots i_s}$ reads
\begin{eqnarray}
&& h = \nonumber \\
&& \sum_{p = 0}^{n}
\sum_{ q = 0}^p
a_{p,q} \ \delta^p   \left(\partial \cdot \partial \cdot\right)^{p-q} \Delta^{n-p+q} Z^{[q]}  . \hspace{1cm} 
\end{eqnarray}
where the coefficients $a_{p,q}$ are determined in the second step.

\subsection{Second Step}
By requesting that a gauge transformation of $Z_{i_1 i_2 \cdots i_s}$ induces a gauge transformation of $h_{i_1 \cdots i_s}$, the coefficients $a_{p,q}$ are found to be given up to an overall multiplicative  constant $a$ by
\begin{eqnarray}
&& a_{p,q} =
\left(-\right)^q
2^{-p} \nonumber \\
&&\frac{n!\left(2n - p\right)!\left(2n+ 1\right)!!}{q!\left(p - q\right)!\left(n - p\right)!\left(2n\right)! \left(2n - 2p +1\right)!!} a .
\end{eqnarray}
The computation is again somewhat fastidious but conceptually straightforward and left to the reader.  

\subsection{Third Step}
Finally, we fix the remaining coefficient $a$ by imposing that $G[h[Z]] = D[Z]$.  This is most conveniently done in the transverse, traceless gauge for $Z$
\be
\partial^{i_1} Z_{i_1 \cdots i_s} = 0, \; \; \; Z^{i_1}_{\; \; \; i_1 i_3 \cdots i_s} = 0
\ee
which is again permissible.  In that gauge, the Cotton tensor reduces to 
\begin{eqnarray}
D\left[ Z \right] &=&
\left(\epsilon \cdot \partial\cdot\right) \Delta^{2n} Z.
\end{eqnarray}
while the Einstein tensor of $h[Z]$ becomes
\begin{eqnarray}
G\left[ h\left[ Z \right] \right] &=&
\left(- \right)^n a \ \epsilon \cdot \partial\cdot \Delta^{2n} Z.
\end{eqnarray}This leads to
\begin{eqnarray}
a &=& \left(-\right)^n.
\end{eqnarray}and completes the determination of $h$ in terms of its prepotential $Z$.

\end{document}